# Preprint WebVRGIS Based Traffic Analysis and Visualization System

*Xiaoming Li[1,2,3], Zhihan Lv[3,*], Weixi Wang[1,2], Baoyun Zhang[4], Jinxing Hu[3], Ling Yin[3], Shengzhong Feng[3]*

1. Shenzhen Research Center of Digital City Engineering, Shenzhen, China
2. Key Laboratory of Urban Land Resources Monitoring and Simulation,
Ministry of Land and Resources, Shenzhen, China
3. Shenzhen Institute of Advanced Technology(SIAT), Chinese Academy of Science, Shenzhen, China
4. Jining Institute of Advanced Technology Chinese Academy of Sicences, Jining, China
* Corresponding Author. Email: lvzhihan@gmail.com. Tel: +86 13791802272

**Abstract**

This is the preprint version of our paper on Advances in Engineering Software. With several characteristics, such as large scale, diverse predictability and timeliness, the city traffic data falls in the range of definition of Big Data. A Virtual Reality GIS based traffic analysis and visualization system is proposed as a promising and inspiring approach to manage and develop traffic big data. In addition to the basic GIS interaction functions, the proposed system also includes some intelligent visual analysis and forecasting functions. The passenger flow forecasting algorithm is introduced in detail.

## 1. Introduction

As an important part of urban transportation system, the urban passenger terminal is an important joint-point for urban internal traffic and external traffic [1]. Integrating transportation routes including many highways and urban roads, multiple means of transportation, it has necessary service function and control equipment, and also has the comprehensive infrastructure which provides places for urban internal and external traffic transition and integrates traffic, commerce, and leisure. The urban passenger terminal often gives a comprehensive consideration to urban external highway passenger transportation and urban public traffic (track, bus, and taxi, etc.), private transportation, as well as railway, aviation, and other external passenger transportation to establish an organic passenger transportation and bring important benefits to urban development [2].

With the evolution and outward extension of urban spatial structure, the lines of public transportation are continuously expanded and extended; meanwhile, the means of transportation has also been developed in a diversified way. The reasonable planning and design and efficient management of urban passenger terminal are the important link to improve urban public transportation system, solve residents' transfer, and improve the service quality and operation benefits of public transportation [3].

Nowadays, there is an increasing interest in using Virtual Reality Geographical Information System (VRGIS), which can obtain the landscape geospatial data dynamically, and perform rich visual 3D analysis, calculations, managements based on Geographical Information System (GIS) data [4] [5] [6] [7] [8]. As a medium composed of interactive computer simulations that sense the participant's position and actions and replace or augment the feedback to one or more senses, virtual reality gives the feeling of being mentally immersed or present in the simulation (a virtual world) [9]. With several characteristics, such as large scale, diverse predictability and timeliness, traffic data falls in the range of definition of Big Data [10]. In addition to the spatial data integration, new user interfaces for geo-databases is also expected [11]. Therefore, the management and development of traffic big data with virtual reality technology is a promising and inspiring approach.

The urban comprehensive passenger transportation hub is a key node of the urban transportation network. With various forms of traffic flows (for example, rail, bus, taxi, long-distance passenger transportation) connected with the hub and complicated forms of transfer among them, the passenger flow transportation efficiency will affect the whole transportation network. In full consideration of the height of the entire hub, the trend of information technology development and application and other aspects, it is pretty necessary to research the real-time state information of the urban comprehensive passenger transportation hub. First, for the daily management of transportation hub, on one hand, the real-time publication of accessibility of all types of passenger flows plays an important role in improving the hub's service quality and optimizing the whole transportation system; on the other hand, researching the residents' long-term traveling needs and the spatial and temporal distribution of passenger flows can help us to grasp the law of passenger flows and estimate the passenger flows, so as to take administrative measures ahead of time. Second, for the planning of transportation hub in Shenzhen, evaluating the service scope and serviceability of the existing transportation hubs is the significant basis for future planning and construction of transportation hubs. Finally, for the technical level, the integration of diversified real-time dynamic transportation information is the inevitable trend. The storage, analysis and 3D visualization of the real-time dynamic transportation information for single nodes in the transportation network and GIS-Transport and other technical trials help to provide technical references for using the diversified real-time dynamic transportation information during the integrated construction of municipal services. Some previous work have inspired our work [12] [13].

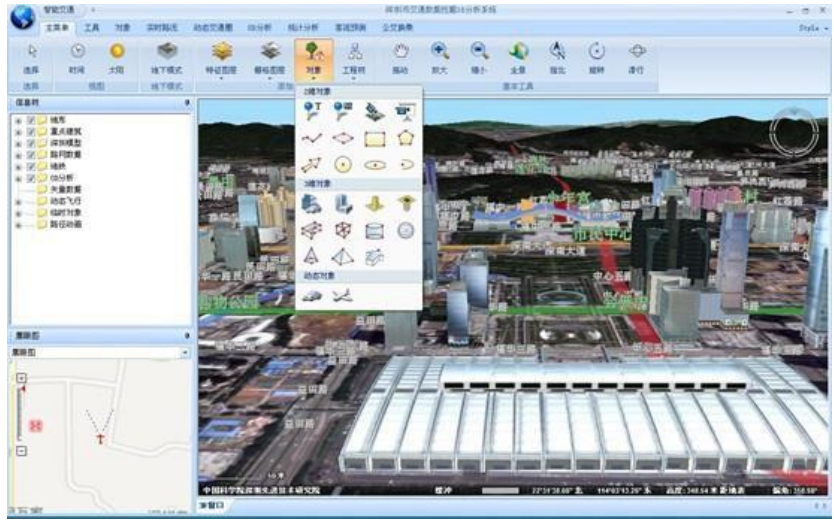

Figure 1: The UI of the virtual reality geographic information system.

## 2. System Division

Currently, there is a lack of integrated analysis and visual display of multiple real-time dynamic traffic information, and also the lack of deep research and application examples on this basis. This research takes Shenzhen Futian Comprehensive Transportation Junction as the case, and makes use of continuous multiple real-time dynamic traffic information (currently using taxi data, the database also includes card swiping data of public transport, and long-distance passenger traffic information, etc.) to monitor out monitoring and analyze the on spatial and temporal distribution of passenger flow under different means of transportation and service capacity of junction from multi-dimensional space-time perspectives such as different period and special period.

Virtual environments have proven to significantly improve public understanding of 3D planning data [14]. To share the information resources of all departments and the dynamic tracking for the geospatial information of population and companies, an integrated information system of social services is constructed. The use of virtual reality as visual means has changed the traditional image of the city [15].

3D visualization of city building is to take inventory and display various types of object data management and resource data within the community area which is so-called virtual community [16], thus helping the social service agencies grasp the work base with the support of holographic foundation library, such as population and companies, houses, events and urban component and other factors. Based on rich Internet application(RIA) technology [17], holographic data query for every household of the specific community can be achieved respectively by building 3D virtual house and associating it with relevant data.

The geographic statistical analysis is to assist management decision-making and conduct data analysis. 2D statistical analysis visualization is overlapped with a white background on 3D virtual reality environment, since it's more intuitive and a cognitively less demanding display system, which lessens the cognitive workload of the user [18]. The innovations this work include, real-time dynamic comprehensive transportation data mining, 3D GIS analysis and release as for transportation junction; comprehensive assessment on service scope of junction by analyzing the long-term dynamic traffic data; comprehensive analysis on actual travel time and then comprehensive assessment on service capacity of public transportation (accessibility and reliability).

Generally, the system can be divided into seven modules, respectively common tool module, real-time traffic status module, dynamic traffic circle module, Origin and Destination (OD)analysis module, passenger flow forecasting module, statistical analysis module and bus transfer module, as shown in Figure 1. The proposed system is based on WebVRGIS [19] and WebVR engine [20].

Basic functions: Switch the mouse into the dragging mode and realize the dragging of 3D earth in any direction according to the mouse's moving direction; then, click to recover. Apart from map magnification and diminishing to display operations, it is also feasible to realize the flight of the systematic map to the panorama location displaying the map of Shenzhen through the panoramic function. The operating ways of map sliding and rotating are also provided. The compass function can indicate the current direction; if you click the compass button, the map will automatically calibrate its direction to the due north direction upwards.

To state the property of road condition, text, picture or video labels can be added on the surface or above the terrain. Besides, 2D lines, 2D faces, 3D models and vector data can also be added on the land surface.

The real-time road condition function can load the information on road conditions in real time and display it from the panoramic angle, as shown in Figure 2. The accessibility analysis function can analyze and display the accessible scope of Futian Transportation Hub in the selected time in the panoramic form, as shown in Figure 3.

Transportation OD analysis: to analyze the number of vehicles and metro trains from the hub to all regional centers according to the data of taxies and metro going in and out of Futian Transportation Hub. The system will display the taxies from Futian Hub to all the regional centers with rays with different degrees of thickness and display the number of trains to all the metro stations at all the locations of metro stations with different cylinders, shown on the left and right of Figure 4.

The passenger flow forecasting function: to analyze and display the daily and monthly taxi passenger flows based on the historical statistical data and through the passenger flow forecasting analysis method. The two different forms of bar graph and curve graph are used to display the daily and monthly passenger flow volume by taxi from Futian Hub, as displayed in Figure 5.

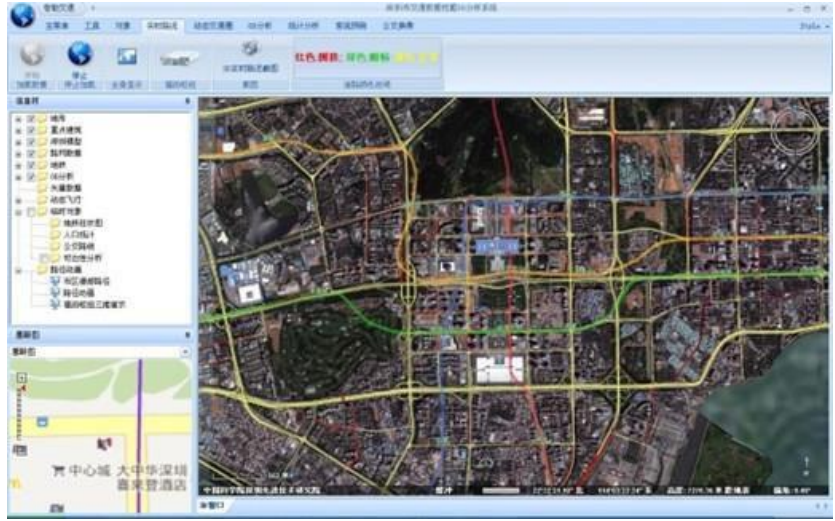

Figure 2: The real-time road condition.

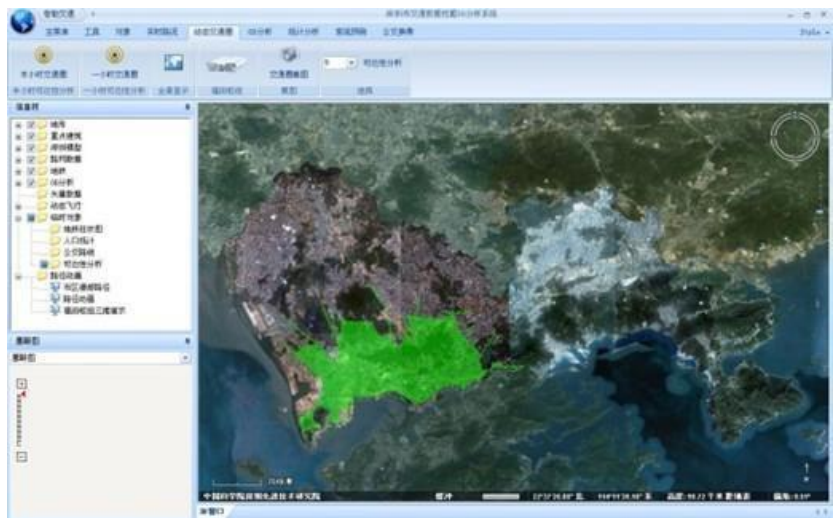

Figure 3: The accessibility analysis.

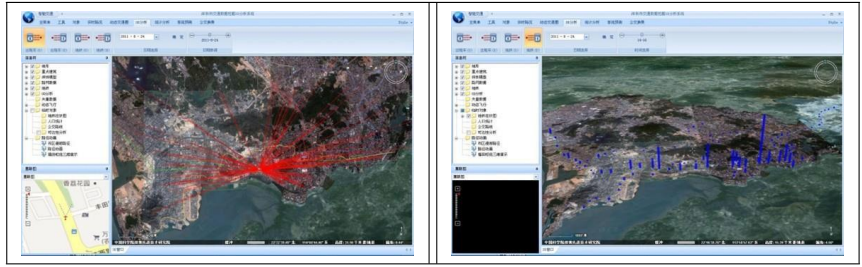

Figure 4: Left: Transportation OD analysis for taxi; Right: Transportation OD analysis for subway.

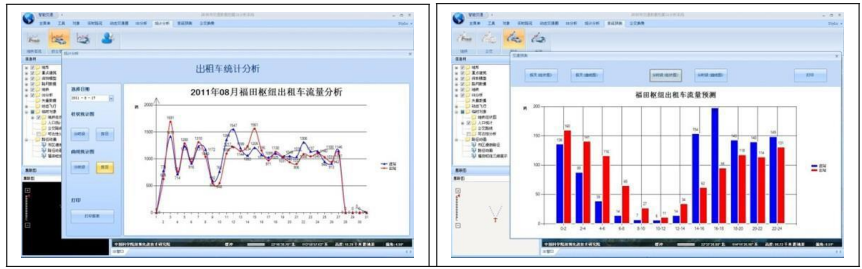

Figure 5: Left: Passenger flow analysis by bar graph; Right: Passenger flow analysis by curve graph.

The bus transfer function is based on the bus route data in Shenzhen City and combines the bus transfer algorithm. The bus transfer query and three-dimensional transfer functions are provided. By entering the original station and terminal station and clicking the Query button, the list will show the bus number to be taken; by clicking the More button, the detailed transfer route will be displayed; by clicking the transfer route, the system will display, on the 3D map, all the stations and the transfer stations where the bus will pass by will appear, as shown in Figure 6.

## 3. Spatial Distribution and Forecast of Taxi Passenger Flow

The taxi passenger flow distribution can directly reflect the condition of ground transportation system of a city at most, so it is a fast and effective way to solve the complicated urban transportation problem by using the spatial and temporal distribution of taxi passenger flow. OD spatial and temporal distribution, which is used to describe the distribution of transportation volumes among all regions of a city and takes the traffic zone as its unit, is a significant component of traffic planning model. The division of traffic zones is a good way to analyze the urban transportation network, because a traffic zone has the similar traffic features and strong traffic relevance. The complete urban transportation system is divided into a lot of traffic zones by the transportation flows, road network layout and other features.

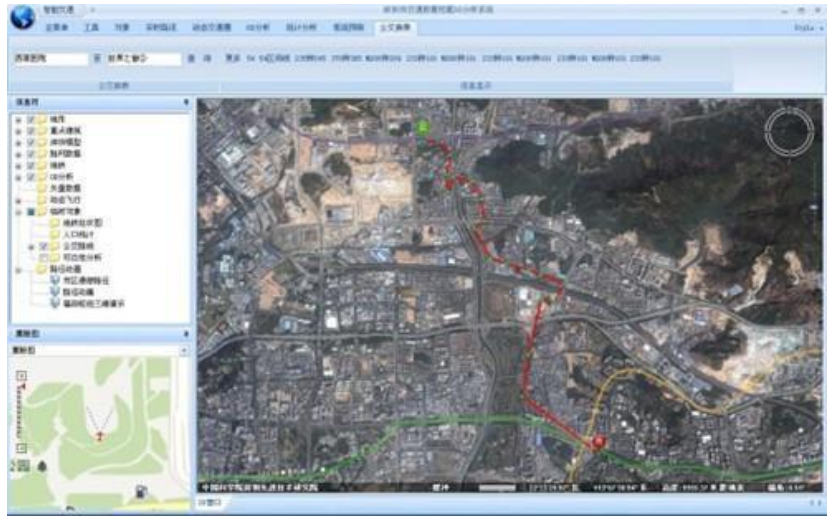

Figure 6: Bus transfer function.

In this research the traffic zones are divided based on the distribution of road networks at all levels in Shenzhen, an OD matrix calculation is conducted according to the taxi probe vehicle GPS data of Shenzhen to offer the OD spatial and temporal distribution of the taxies with passengers and deduce the OD spatial and temporal distribution about transportations between the city and Futian Transportation Hub.

## 4. Analysis on Taxi Passenger Flow Volume

The data used in this forecast mainly include the real-time GPS data of vehicle. The original data sheet mainly saves the data gathered by the GPS terminal installed in the taxi, including license plate number, time point of data collection, longitude, latitude, vehicle state, vehicle speed, driving direction, empty and load state, etc. Due to the huge data volume every day, in consideration of such aspects as shortening the data query time and improving the overall operational performance, we have sorted the original data.

The following statistics in Figure 7 indicate the day-by-day variation in the number of taxies entering and exiting from the station.

The horizontal axis represents the time, and the vertical axis represents the vehicle flow volume. The data include those collected for 27 days in August, 14 days for September and 24 days for October; the data for other days between August and October are absent. According to the analysis on the data collected for 65 days during the three months, we find that the factors affecting the taxi flow volume are complex. Among the statistics of taxi flow volume, the flow volumes of taxies entering and exiting from the station are basically the same. By observing the data, it can be learned that over 1500 vehicles entered the station on August 12, September 30 and October 1 respectively, and that over 1500 vehicles exited from the station on August 3, August 15 and October 1; according to the investigation, it was learned that such phenomenon was caused by the 26th World University Games held in Shenzhen City from August 12 to August 23, in which period the traffic restriction, flow division at weekends and other measures were implemented to avoid traffic congestion; therefore, the variation trend of taxi flow volume in this period was not the same as usual; this was why a peak flow volume of taxies entering the station occurred on August 12 when the opening ceremony was held, and a peak flow volume of taxies exiting the station occurred on August 15. A before-festival peak passenger flow volume occurred in the National Day golden week from October 1 to October 7, and a valley formed quickly after the end of the golden week. Excluding the two periods of time, no any obvious difference was found in the passenger flow volumes of weekdays and weekends, indicating that the crowds served by taxies are different from those served by other public transportations, and that the trip purposes of passenger crowds are particular.

The analysis above indicates that there is no obvious variance for the taxi passenger flow volume at Futian Transportation Hub Station has between weekdays and weekends; but there is the rule that the flow volume entering the station is large in the morning and the flow volume exiting the station is also large in the evening; indicating that the main crowd served may be the long-distance passenger crowd; they start off in the morning and come back in the evening by taking the long-distance bus.

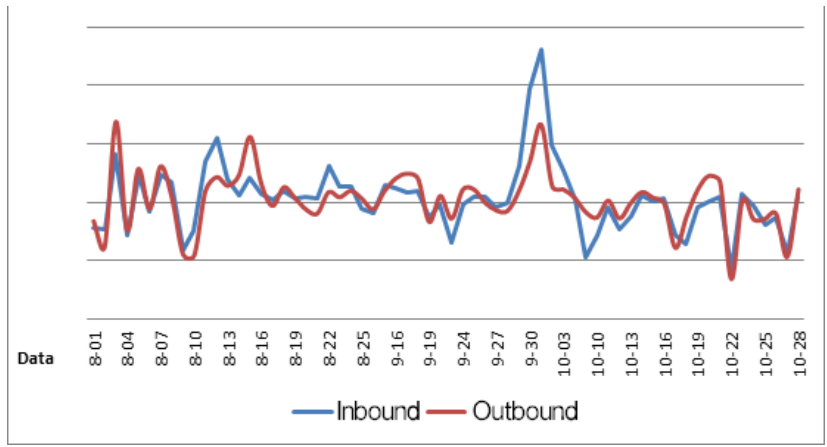

Figure 7: Flow Volume of Taxies Entering and Exiting the Futian Transportation Hub Station Between August and October. Y-axis means the amount of the taxis.

## 5. Spatial Distribution of Taxi Passenger Flow

By adopting the regression forecast model, our research forecasts the passenger flow of taxi. While predicting the passenger flow, there possibly many independent variables affecting the result of dependent variable (passenger flow). As a matter of fact, it is only permitted to find out some independent variables having important influence on the dependent variable, but neglect other independent variables. In concrete application, it needs to screen out some major independent variables that may affect the result of dependent variable, such a screening progress will apply the variance analysis theory. In statistics, condition controlled in the experiment is factor. The state of factor is processing or level. According to the experimental result is affected by one or several factors, it can be divided into single-factor experiment and multi-factor experiment. The sample of passenger flow varies at different stages every day. Therefore, it can be concluded that date and time frame are two of many variable factors affecting the passenger flow. By carrying out two-factor analysis of variance, it can get that whether these two factors are important one affecting the passenger flow, further to consider them as input variables in the subsequent predictive analysis.

According to the transportation network, we divide the six districts of Shenzhen City into 228 traffic zones; Futian District and Luohu District have the largest number of traffic zones, because of their dense population, economic prosperity and complicated industrial structures; the division of traffic zones can basically reflect the urban population distribution, economic characteristics, industrial structures, traffic flow and other features. Figure 8 left and Figure 8 right indicate the distribution of daily taxi passenger sources and daily passenger flow respectively at Futian Transportation Hub Station. These figures have revealed the major distribution conditions of the crowds served by taxies; the passenger flow distribution is basically the same as the passenger source distribution, indicating that the main crowds served by taxies are distributed in those zones with dense pollution and prosperous economy. Besides, the range of taxi service extends 9km towards the east and the west. According to the spatial and temporal distribution characteristics of taxi passenger crowds, it is found that the crowds arrive at Futian Transportation Hub Station by taxi with the main purpose of taking the long-distance bus, and that the crowds taking taxies from the Futian Transportation Hub Station arrived the station mainly by long-distance bus; the main foundation of a taxi is for transferring between long-distance passenger flow and urban passenger flow.

## 6. Taxi passenger flow forecast

The passenger flow forecast refers to the index reflecting the demand characteristics of traffic passenger flow by forecasting the cross sectional flow of urban transport lines within certain period and inter-station OD. Upon planning the traffic network, the passenger flow analysis result of different traffic network schemes is the main content based on which the line network is selected.

People's travel demand is the unity of randomness and regularity [21], and it is reflected in large quantity of statistical data. Therefore, it is difficult to accurately forecast the passenger flow in one day; however, the forecast can be made within certain confidence, which is also the starting point of forecasting.

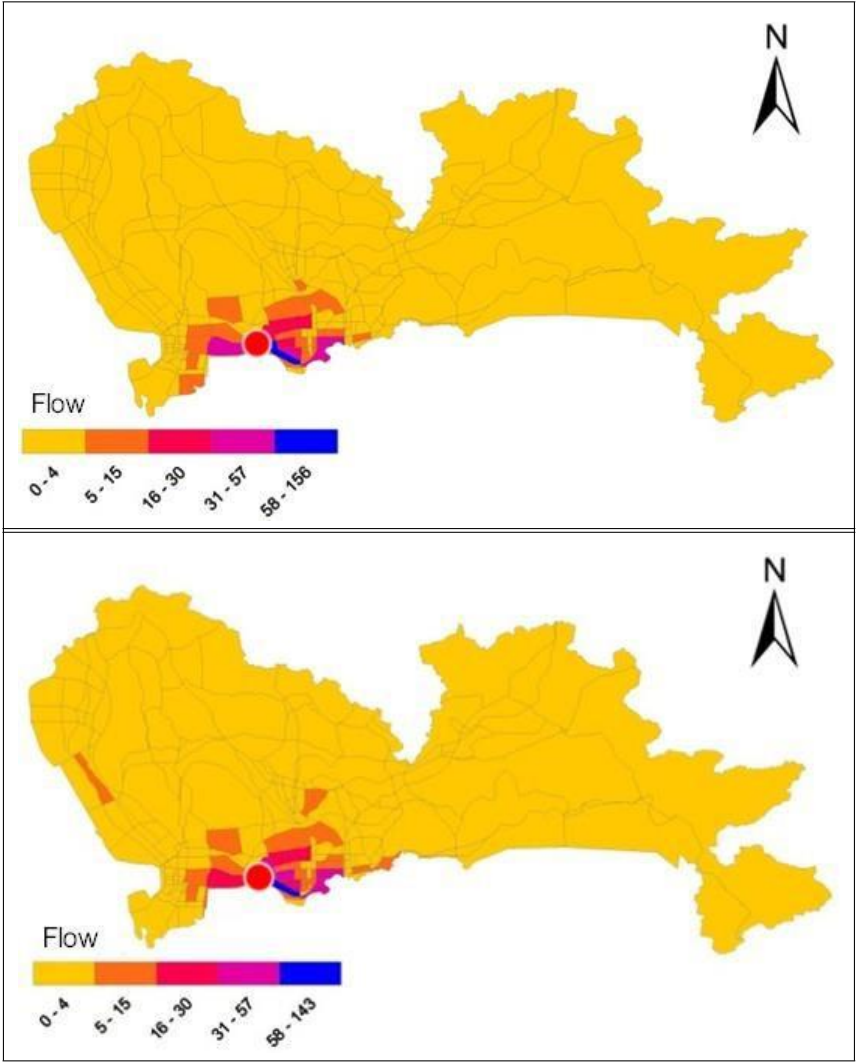

Figure 8: Left: Distribution of Daily Taxi Passenger Flow Arriving at Futian; Right: Distribution of Daily Taxi Passenger Flow Departing from Futian.

There are many kinds of traffic passenger flow forecast models, and the common models include regression forecast model and time series prediction model [22] [23]. In the observation period of this research, the passenger flow shows strong regularity and stability without long-term change trend; therefore, regression forecasting model is employed in this paper to forecast the passenger flow. While we forecast the passenger flow, there may be many independent variables influencing the result of dependent variable (passenger flow), but the actual situation is that it is only allowed to find out several independent variables which have important influence on the dependent variable and ignore other independent variables. In specific application, it is required to screen out some main independent variables which influence the result of dependent variable for research and analysis, and the analysis of variance theory is applied in this screen-out process.

The analysis of variance is a kind of statistical analysis method in which the analysis and processing is made for significance of difference in mean values of some sets of experimental data. The passenger flow samples are different in different periods everyday; therefore, it is able to know that the date and period are 2 variables which influence passenger flow; through dual-factor analysis of variance, it is able to determine whether these two factors are important factors which influence passenger flow and then regard them as input variables in later forecasting analysis if so.

Due to difference in inbound and outbound passenger flow, this research carries out forecasting modeling and model validation on inbound and outbound passenger flow respectively.

For outbound traffic, predictive modeling and model validation contain the following steps:

1. Outbound traffics of twenty-six days with normal data at each time period from August to October are selected as modeling samples.

2. This study divides time period into 12 parts, and makes time period variable become a categorical independent variable with 12 types, and thus 11 dependent variables are generated, as follows:

$Flow\text{-}in = b_1 t_1 + b_2 t_2 + b_3 t_3 + b_4 t_4 + b_5 t_5 + b_6 t_6 + b_7 t_7 + b_8 t_8 + b_9 t_9 + b_{10} t_{10} + b_{11} t_{11} + b_{12}$ (1)

Table 1: Regression modeling results of outbound traffics regression statistical results

| Regression | |
|---|---|
| Multiple R | 0.906573 |
| R Square | 0.821875 |
| Adjusted R Square | 0.815321 |
| Standard Error | 22.81944 |
| Observed Value | 311 |

Table 2: Regression modeling results of outbound traffics variance analysis results

| | df | SS | MS | F | Significance F |
|---|---|---|---|---|---|
| Regression Analysis | 11 | 718390.5 | 65308.23 | 125.4175 | 5.1E-105 |
| Residual Error | 299 | 155697.3 | 520.7268 | | |
| Total | 310 | 874087.8 | | | |

Where, $t_1, t_2, \ldots t_{11}$ are dependent variables of 11 time periods. When $t_1 = 1, t_2 = t_3 = \ldots = t_{11} = 0$, dependent variables denote outbound traffics of the first time period; when $t_2 = 1, t_1 = t_3 = \ldots = t_{11} = 0$, dependent variables denote outbound traffics of the second time period; reason by analogy; when $t_1 = t_2 = t_3 = \ldots = t_{11} = 0$, dependent variables denote outbound traffics of the twelve time period. $b_i$ is independent variable parameters and constant terms of these four models respectively.

3. By modeling with sample data in Table 1, results of predictive model of outbound taxi are as follows:

According to Table 1, predictive model of outbound traffics has Adjusted R Square=81.53%, i.e. this model can explain 81.53% of sample data; the entire model has statistical significance at level $\alpha = 0.05$; all parameters of the model have p-value smaller than 0.05, which means that all of these parameters have statistical significance at level $\alpha = 0.05$.

To sum up, predictive models of outbound traffics have good imitative effects of sample data and are expressed as follows:

Table 3: Regression modeling results of outbound traffics model coefficient results

|  | Coefficients | Standard Error | t Stat | P-value |
|---|---|---|---|---|
| Intercept | 54.07692 | 4.47526 | 12.08353 | 1.21E-27 |
| 1 | -28.4369 | 6.39195 | -4.44887 | 1.22E-05 |
| 0 | -46.2308 | 6.328973 | -7.30462 | 2.54E-12 |
| 0 | -21.8462 | 6.328973 | -3.45177 | 0.000637 |
| 0 | 2.384615 | 6.328973 | 0.376778 | 0.706606 |
| 0 | 36.11538 | 6.328973 | 5.706357 | 2.78E-08 |
| 0 | 65.73077 | 6.328973 | 10.38569 | 8.88E-22 |
| 0 | 69.07692 | 6.328973 | 10.9144 | 1.45E-23 |
| 0 | 82.57692 | 6.328973 | 13.04744 | 4.18E-31 |
| 0 | 98.38462 | 6.328973 | 15.54511 | 2.39E-40 |
| 0 | 89.61538 | 6.328973 | 14.15955 | 3.49E-35 |
| 0 | 52.26923 | 6.328973 | 8.258722 | 4.81E-15 |

Table 4: Validation results of predictive model of taxi outbound traffics at different time periods

| Maximum Error | Minimum Error | Mean Error |
|---|---|---|
| 125.32% | 1.23% | 23.46% |

$$Flow\_out = -28.44t_1 - 46.23t_2 - 21.85t_3 + 2.38t_4 + 36.16t_5 + 65.73t_6 + 69.08t_7 + 82.58t_8 + 98.38t_9 + 89.62t_{10} + 52.27t_{11} + 54.08 \quad (2)$$

Outbound traffics at time periods with normal data within two days of October are selected as model validation samples. Validate prediction formula with validation sam- ples and calculate the accuracy of each independent predicted value with mean absolute percentage error (MAPE): $MAPE = 100\% \times |predicted value - actual value|/actual value$. Validation results are as Table 4.

As can be known from Table 3, the mean error of predictive model of taxi outbound traffics at different time periods is 23.46%, and thus this model has good validation effects.

## 7. Conclusion

In this research, Futian Transportation Junction is taken as research objective, and the spatial and temporal distribution rules of passenger flow and service scope of junction of all means of transportation are found by carrying out statistical analysis on long-term daily passenger flow and daily time-phased passenger flow in Futian Transportation Junction according to the data such as Shenzhen TransCard data, taxi and floating car data, and long-distance passenger transportation data, and a short-term forecast is made on various kinds of passenger flow. Furthermore, the key analysis is made on abnormal traffic condition in the period of Universiade and National Day holidays. It is found from the research that Futian Transportation Junction currently has good operation condition, and the passenger flow of various means of transportation doesn't reach the designed upper limit of passenger flow even in peak of passenger flow in holidays, without obvious transfer and congestion. Within short term, various means of transportation have stable passenger flow, and the time-phased passenger prediction model is reliable. The main service group of junction is distributed in west of Nanshan District and Futian District, as well as west of Luohu District and the scope which radiates about 9km based on the junction.

Then, the travel time is used as index to carry out a series of researches and analyses on accessibility of public transportation of Futian Transportation Junction. It is found from the research that the junction has good accessibility. Then, the actual fluctuation of travel time is taken as index to analyze the reliability of junction network. It is found that the overall reliability is high, and there is only poor reliability at the peak travel time at Huaqiang North station and Bao'an center station; as for some business centers and hot working area, there is poor reliability at the peak travel time; as for places with strong entertainment such as Overseas Chinese Town and Honey Lake, there is a poor reliability at travel time in holidays.

In addition, the 3D case of Shenzhen proves that 3D city visualization and analysis system is a useful tool for the social service agencies and citizens for browsing and analyzing city big data directly, and is agreed upon as being both immediately useful and generally extensible for future applications.

## 8. Future Work

Through long-term monitoring and analysis, the long-term passenger flow forecast is established under the condition of combing with economic development and urban planning. Our future analysis will be made on population travel behavior, taxi route, degree of influence on public bus, and road travelling speed under special weather conditions (such as rainstorm, typhoon, and heavy fog). The deeper data mining will be made, such as emergency evacuation aided decision support, monitoring and forecasting on large-scale group event, assisting crowd and vehicle evacuation under emergency. In addition to city, ocean data will be integrated into the proposed system [24] [25], in which we will employ spatiotemporal visualization as the representative approach [26] [27]. Some novel human-computer-interaction approaches [28] [29] [30] [31] [32] [33], enhanced vision feedback technologies [34] [35] [36] [37] [38], network technologies [39] [40] [41] [42], bigdata [43] [44] [45] [46] and some optimization approaches [47] [48] [49] [50] [51] are considered to be integrated in our future work.

## Acknowledgments

The authors are thankful to the National Natural Science Fund for the Youth of China (41301439), LIESMARS Open Found(11I01) , Shenzhen Scientific & Research Development Fund (JC201105190951A, JC201005270331A, CXZZ20130321092415392), and Electricity 863 project(SS2015AA050201).

cageo.2010.04.016.
URL http://dx.doi.org/10.1016/j.cageo.2010.04.016

[12] C. Zhong, S. M. Arisona, X. Huang, M. Batty, G. Schmitt, Detecting the dynamics of urban structure through spatial network analysis, International Journal of Geographical Information Science 28 (11) (2014) 2178–2199.

[13] C. Zhong, X. Huang, S. M. Arisona, G. Schmitt, M. Batty, Inferring building functions from a probabilistic model using public transportation data, Computers, Environment and Urban Systems 48 (2014) 124–137.

[14] E. Chow, A. Hammad, P. Gauthier, Multi-touch screens for navigating 3d virtual environments in participatory urban planning, in: CHI '11 Extended Abstracts on Human Factors in Computing Systems, CHI EA '11, ACM, New York, NY, USA, 2011, pp. 2395–2400. doi:10.1145/1979742.1979852.
URL http://doi.acm.org/10.1145/1979742.1979852

[15] C. DiSalvo, J. Vertesi, Imaging the city: Exploring the practices and technologies of representing the urban environment in hci, in: CHI '07 Extended Abstracts on Human Factors in Computing Systems, CHI EA '07, ACM, New York, NY, USA, 2007, pp. 2829–2832. doi:10.1145/1240866.1241088.
URL http://doi.acm.org/10.1145/1240866.1241088

[16] Z. Lu, S. U. Rehman, G. Chen, Webvrgis: Webgis based interactive online 3d virtual community, in: Virtual Reality and Visualization (ICVRV), 2013 International Conference on, IEEE, 2013, pp. 94–99.

[17] M. Zhang, Z. Lv, X. Zhang, G. Chen, K. Zhang, Research and application of the 3d virtual community based on webvr and ria, Computer and Information Science 2 (1) (2009) P84.

[18] T. Porathe, J. Prison, Design of human-map system interaction, in: CHI '08 Extended Abstracts on Human Factors in Computing Systems, CHI EA '08, ACM, New York, NY, USA, 2008, pp. 2859–2864. doi:10.1145/1358628.